# Open Problems in Relational Quantum Mechanics

Federico Laudisa

*Department of Human Sciences, University of Milan-Bicocca*

Piazza dell'Ateneo Nuovo 1, 20126 Milan – Italy

federico.laudisa@unimib.it

**Abstract**

The Rovelli relational interpretation of quantum mechanics (RQM) is based on the assumption according to which the notion of observer-independent state of a physical system is to be rejected. In RQM the primary target of the theory is the analysis of the whole network of relations that may establish among quantum subsystems, and the shift to a relational perspective is supposed to address in a satisfactory way the general problem of the interpretation of quantum mechanics. Here I argue – mainly through an analysis of the so-called *third person problem* – that it is far from clear what a relativization of states to observers exactly achieves and in what sense such an approach really advances our understanding of the peculiar features of quantum phenomena: therefore, in this respect, RQM still faces open problems.

## 1   Introduction

The peculiar way in which some apparently 'weird' relations among subsystems emerge in quantum mechanics (QM) has been the *locus* of a long-standing debate in the foundations of the theory: as is well-known, the highly non-classical composition of physical systems into more complex systems in QM implies the existence of entangled states and, as a consequence, the existence of non-classical correlations among subsystems[1]. The so-called relational interpretation

---

[1] Consider for instance the debate that already more than thirty years ago was fueled by the Teller *relational holism* (Teller 1986, 1989; see Morganti 2009 for a general assessment) or the issue of the priority of the structure of relations over individuals in the debate on the structural realism (Ladyman 2014).

of quantum mechanics (RQM from now on) – proposed in the late Nineties by Carlo Rovelli – is based on the assumption according to which the primary object of the theory is exactly the analysis of the whole network of relations that may establish among quantum subsystems. As a consequence of this theoretical choice, the starting point of RQM is the intentional rejection of the notion of *observer-independent state of a physical system*, also on the basis of an intuitive resemblance with the emergence of special relativity and the related rejection of the assumption of an absolute time, encoded into the Galileian transformations:

The Lorentz transformations were perceived as unreasonable, even inconsistent, before 1905. […] Einstein discovered the reason for the unease: the implicit use of a concept (observer-independent time) inappropriate to describe reality or, equivalently, a common assumption about reality (simultaneity is observer-independent) that was physically incorrect. The unease with Lorentz transformations derived from a conceptual scheme in which an *incorrect notion*, absolute simultaneity, was assumed, yielding many sorts of paradoxical situations. I suspect that the "paradoxical" situations associated with quantum mechanics may derive from some analogous *incorrect notion* that we still employ in thinking about quantum mechanics […] Such a notion, I maintain, is the notion of absolute, or observer-independent, state of a system. (Rovelli 1997, pp. 196-197, emphasis in the original)

In addition to this analogy, a more robust ground for such a starting point comes originally from the quantum gravity research program: namely, RQM is supposed to be the formulation of QM that the need for a plausible theoretical framework for quantum gravity forces on us, under the assumption that the main lesson we learn from General Relativity (GR) lies in its relational description of the motion of all dynamical entities. In a section of his book on the quantum gravity program, a section devoted to RQM, Rovelli emphasizes this motivation:

The main idea underlying GR is the relational interpretation of localization: objects are not located in spacetime. They are located with respect to one another.  In this section, I have observed that the lesson of QM is that quantum events and states of systems are relational: they make sense only with respect to another system. Thus both GR and QM are characterized by a form of relationalism. Is there a connection between these two forms of relationalism? (Rovelli 2004, p. 220)

But there is more: according to Rovelli, the shift to a relational perspective can be helpful even for QM *alone*, in that it allows one to address in a satisfactory way the general problem of the interpretation of the theory. In fact, Rovelli takes an ambitious step in claiming that (i) by dropping the notion of observer-independent state "quantum mechanics makes much more sense", and that (ii) the above conclusion "derives from the observation that the experimental evidence at the basis of quantum mechanics forces us to accept that distinct observers give different descriptions of the same events." (Rovelli 1996, pp. 1638-9). Moreover, in RQM locality would be recovered: "in the context of this interpretation, it is not necessary to abandon locality in order to account for EPR correlations. From the relational perspective, the apparent «quantum non-locality» is a mistaken illusion caused by the error of disregarding the quantum nature of *all* physical system" (Smerlak, Rovelli 2007, p. 428). By a relational viewpoint, this recovery is also taken to contribute to a synthesis of QM with GR, given the fundamentally local nature of GR itself.

In the present paper I will argue that both claims (i) and (ii) are controversial, and that this brings to bear on the locality issue within RQM: as such, RQM still faces open problems, since it is far from clear what a relativization of states to observers exactly achieves and in what sense such an approach really advances our understanding of the peculiar features of quantum phenomena. In section 2 I will give a general overview of RQM, whereas in section 3 I will focus on the reasons why the alleged relativity of observers' state descriptions that is at the heart of RQM seem to depend on a fundamental ambiguity in the analysis of the quantum-mechanical measurement process in the standard formalism. Section 4 will be devoted to a discussion of the status of the EPR argument in RQM and, finally, a general conclusion will be drawn in section 5.

## 2 Relational quantum mechanics: an overview

A general outline of RQM[2] can be presented starting from a key remark concerning ordinary QM: in its standard formulation, the theory appears to suffer from an indeterminacy problem in describing the main process by which an experimenter is supposed to extract physically relevant information in an experimental context, namely the infamous measurement process. This problem is not a practical one – in light of the impressive predictive and experimental success of the theory – but a foundational one: the linearity of the theory implies superpositions of states that fail to be reflected in the recordings of the measurement outcomes, that in fact turn out to be always determinate. This unsatisfactory state of affairs – which was emphasized already by von Neumann in his classic treatise on the mathematical foundations of QM (von Neumann 1932, ch. 6) and which is probably the most serious problem that the contenders in the interpretational debate have attempted to solve – might be addressed according to RQM by changing the very notion of state of a quantum system:

[The relational interpretations of QM] are based on the idea that the theory should be understood as an account of the way distinct physical systems affect each other when they interact – and not the way physical systems 'are'. This account exhausts all that can be said about the physical world. The physical world must be described as a net of interacting components, where there is no meaning to 'the state of an isolated system'. The state of a physical system is the net of the relations it entertains with the surrounding systems. The physical structure of the world is identified as this net of relationships. More precisely, the way out suggested by the relational interpretations is that the values of the variables of a physical system $S$ […] are relational. That is, they do not express a property of the system alone, but rather refer to the relation between two systems. (Rovelli 2005, p. 115)

In RQM the relativization of states is claimed to solve "the apparent contradiction between the two statements that a variable has or does not have a value by indexing the statements with the different systems with which the system in question interacts": as a consequence, "the unique

---

[2] The outline of RQM given here is rather sketchy and instrumental to the critical points I wish to discuss in the central sections of the paper. For a wider presentation, sensitive also to the metaphysical background and implications of RQM, see Dorato 2016.

account of the state of the classical world is thus fractured into a multiplicity of accounts, one for each possible 'observing' physical systems". According to Rovelli, "the central idea of RQM is to apply Bohr and Heisenberg's key intuition that «no phenomenon is a phenomenon until is an observed phenomenon» to each observer independently." (Smerlak, Rovelli 2007, p. 429). The sort of ontology underlying RQM is an ontology of *events*, out of which what we ordinarily call 'physical systems' emerge:

In RQM physical reality is taken to be formed by the individual *quantum events* ('facts') through which interacting systems (objects) affect one another. Quantum events are therefore assumed to exist only in interactions and (this is the central point) *the character of each quantum event is only relative to the system involved in the interaction*. In particular, which properties any given system *S* has is only relative to a physical system *A* that interacts with *S* and is affected by these properties. (Smerlak, Rovelli 2007, pp. 429-430).

In RQM the ensuing 'fragmentation' of descriptions, however, need not affect the *completeness* of the theory: "QM is a theory about the physical description of physical systems relative to other systems, and this is a complete description of the world (Rovelli 2005, p. 116).

Before going on, some remarks are in order. First, in the intuition of relativizing states to 'observers', the notion of *observer* is not meant to be 'subjective' in any sense: "By using the word *observer* I do not make any reference to conscious, animate or computing, or in any other manner special, systems. I use the word *observer* in the sense in which it is conventionally used in Galileian relativity when we say that an object has a velocity *with respect to a given observer*. The observer can be any physical object having a definite state of motion." (Rovelli 1996, p. 1641).

Second, RQM differs from a many-world or Everett-like interpretation[3]. In a many-worlds or Everett-like interpretation, the true ontology that the interpretation attempts to explain and defend is the ontology of the wave function of the entire universe: the 'splitting' into relative states, which is inherent to the very formalism of QM, makes all these states to somehow *co-exist*, so that relationalism is a feature of a true *multiplicity* of worlds, branches or whatever (for a recent defense see Vaidman 2016). In RQM, instead, "the quantum events *q*, that is, the actualizations of values of physical quantities, [are] the basic elements of reality and such *q*'s are assumed to be univocal. The relational view avoids the traditional difficulties in taking the *q*'s as univocal simply by noticing that a *q* does not refer to a system, but rather to a *pair* of systems." (Laudisa, Rovelli 2008, emphasis added).

---

[3] As is well known, what is the story that the Everett interpretation exactly tells is a matter of controversy. As Barrett emphasizes: "There has been considerable disagreement over the precise content of his theory and how it was supposed to work." (Barrett 2014).

Third, a crucial role is played by the notion of information as correlation (more on this later): "that «*S* has information about *q*» means that there is a correlation between the variable *q* and the pointer variable […] This result provides a motivation for the use of the expression «information» because information is correlation." (Rovelli 1996, p. 1653). This Rovelli approach to the notion of information has been characterized by van Fraassen as a *transcendental* approach:

Rovelli describes not the world, but the general form of information that one system can have about another – namely, as the assignment of states relative to a given system on the basis of information available to that system:
- there is no implication of possible specific information about what there is which is independent of any point of view, but
- there can be knowledge of the form that any such information, relative to a particular vantage point, must take.

So we have here a *transcendental* point of view. Rovelli offers us this knowledge of the general form, the conditions of possibility. We must take very seriously the fact that as he sees it, quantum mechanics is not a theory about physical states, but about ('about'?) information. The principles he sees at the basis of quantum mechanics are principles constraining the general form that such information can take, not to be assimilated to classical evolution-of-physical-states laws. (van Fraassen 2010, pp. 397-8)[4]

Finally, a remark useful to clearly define the scope of the present paper. In addition to defending the decision to drop the concept of observer-independent values of physical quantities, Rovelli advances a further claim, namely "that quantum mechanics will cease to look puzzling only when we will be able to *derive* the formalism of the theory from a set of simple physical assertions («postulates», «principles») about the world. Therefore, we should not try to append a reasonable interpretation to the quantum mechanics formalism, but rather to derive the formalism from a set of experimentally motivated postulates." (Rovelli 1996, p. 1639). This is a different (and over-ambitious!) issue, that I will not touch in the present investigation[5].

---
[4] On a possibly transcendental view of RQM one *can see* also Bitbol 2008.

## 3   RQM and the "third person" problem

As recalled above, a key starting suggestion at the basis of RQM is the outcome of what Rovelli called originally the "third person problem" [6]. With the symbol $O$ we denote an observer, that at time $t_1$ measures the observable **A** on the system $s$. The measurement of **A** on $s$ by $O$ can be represented via standard assumptions, namely (i) the observable **A** can assume two values (call them 1 and 2), so that $|1>$ and $|2>$ denote respectively the corresponding eigenstates of **A**; (ii) the system $s$ can be prepared in a state $a|1> + b|2>$, with $|a|^2$ and $|b|^2$ being the probabilities attached respectively to values 1 and 2.

If the measurement gives the value 1, then the physical sequence of events $E$ can be represented as follows:

$$\left.\begin{array}{ccc} t_1 & \Rightarrow & t_2 \\ \\ a|1> + b|2> & \Rightarrow & |1> \end{array}\right\} E$$

In terms of the ordinary formulation of quantum mechanics, this is nothing but the outcome of the state reduction process, that drives the initial superposition in one or the other of the two possible values. Now let us consider the same sequence as described by a second observer $P$,

---

[5] This claim of Rovelli recalls one of the aims of the approaches to QM in the tradition of quantum logic and quantum information: for a possible assessment of a reconstruction attempt along the lines suggested by Rovelli see Grinbaum 2007.
[6] See for instance Rovelli 1996, p. 1642-1643, Brown 2009, pp. 682-683, van Fraassen 2010, pp. 396-397, Laloë 2012, pp. 222-223, Smerlak 2017, p. 196.

that 'describes' the measurement of **A** on *s* by *O*: "We assume that *P* uses conventional QM. We also assume that *P does not perform any measurement* on the *s+O* system during the $t_1$ - $t_2$ interval, but that she knows the initial states of both *s* and *O*, and is able to give a quantum mechanical description of the set of events ***E***." (Rovelli 1996, p. 1642). Now the state space of *s+O* is the tensor product $H_s \otimes H_O$: this space includes the states |*O*-ready>, |*O1*>, |*O2*>, where

|*O*-ready> = the state of the observer *O* prior to the measurement

|*O1*> = the state of the observer *O* finding the measurement value 1

|*O2*> = the state of the observer *O* finding the measurement value 2

$$t_1 \quad \Rightarrow \quad t_2$$

$$(a|1> + b|2>) \otimes |O\text{-ready}> \quad \Rightarrow \quad a|1> \otimes |O1> + b|2> \otimes |O2> \quad \Big\} \ \boldsymbol{E'}$$

To sum up: the sequence ***E*** is relative to observer *O*, the sequence ***E′*** is relative to observer *P* and ***E*** ≠ ***E′***: «In the *O* description, the system *s* is in the state |1> and the quantity [**A**] has value 1. According to the *P* description, *s* is not in the state |1> and the hand of the measuring apparatus does not indicate '1'.» (Rovelli 1996, p. 1643).

Several are the problems of this account. First, it appears to turn an indeterminacy into a postulate, by reading 'in positive' such indeterminacy as an indication of the allegedly relational nature of the overall theory. But the difference between ***E*** and ***E′*** is far from a relational flag: it is the heart of a deep problem that the theory is committed to *overcome*. For the 'third person problem' is nothing but another name for the thought experiment known as *the Wigner friend* :

In order to present this argument, it is necessary to follow my description of the observation of a "friend" in somewhat more detail than was done in the example discussed before. Let us assume again that the object has only two states $\psi_1$ and $\psi_2$. If the state is, originally, $\psi_1$, the state of object plus observer will be, after the interaction, $\psi_1 \times \chi_1$; if the state of the object is $\psi_2$, the state of object plus observer will be $\psi_2 \times \chi_2$ after the interaction. The wave functions $\chi_1$ and $\chi_2$ give the state of the observer; in the first case he is in a state which responds to the question "Have you seen a flash?" with "Yes"; in the second state, with "No". There is nothing absurd in this so far. Let us consider now an initial state of the object which is a linear combination $\alpha\, \psi_1 + \beta\, \psi_2$ of the two states $\psi_1$ and $\psi_2$. It then *follows* from the linear nature of the quantum mechanical equations of motion that the state of object plus observer is, after the interaction, $\alpha(\psi_1 \times \chi_1) + \beta(\psi_2 \times \chi_2)$. If I now ask the observer whether he saw a flash, he will with a probability $|\alpha|^2$ say that he did, and in this case the object will also give to me the responses as if it were in the state $\psi_1$. If the observer answers "No" – the probability for this is $|\beta|^2$– the object's responses from then on will correspond to a wave function $\psi_2$. […] All this is quite satisfactory: the theory of measurement, direct or indirect, is logically consistent so long as I maintain my privileged position as ultimate observer. However, if after having completed the whole experiment I ask my friend, "What did you feel about the flash before I asked you?" he will answer, "I told you already, I did [did not] see a flash" as the case may be. In other words, the question whether he did or did not see the flash was already decided in his mind, before I asked him. If we accept this, we are driven to the conclusion that the proper wave function immediately after the interaction of friend and object was already either $\psi_1 \times \chi_1$ or $\psi_2 \times \chi_2$ and not the linear combination $\alpha(\psi_1 \times \chi_1) + \beta(\psi_2 \times \chi_2)$. This is a contradiction, because the state described by the wave function $\alpha(\psi_1 \times \chi_1) + \beta(\psi_2 \times \chi_2)$ describes a state that has properties which neither $\psi_1 \times \chi_1$ or $\psi_2 \times \chi_2$ has. If we substitute for "friend" some simple physical apparatus, such as an atom which may or may not be excited by the light-flash, this difference has observable effects and *there is no doubt that $\alpha(\psi_1 \times \chi_1) + \beta(\psi_2 \times \chi_2)$ describes the properties of the joint system correctly, the assumption that the wave function is either $\psi_1 \times \chi_1$ or $\psi_2 \times \chi_2$ does not*. If the atom is replaced by a conscious being, the wave function $\alpha(\psi_1 \times \chi_1) + \beta(\psi_2 \times \chi_2)$ (which also follows from the linearity of the equations) appears absurd because it implies that my friend was in a state of suspended animation before he answered my question. (Wigner 1967, pp. 179-180).

This somewhat paradoxical situation was devised by Eugene Wigner to support nothing less than his bold claim about QM, that is, the necessity to admit the existence of consciousness in order for the laws of QM to make sense. The radical character of this claim is a counterpart to the coexistence of the <account-with-friend> and the <account-without-friend>, something that Wigner interprets *not* as a sign of any fundamental relationality in the quantum-mechanical description but rather as an absurdity: whether the Wigner friend did or did not see the flash cannot depend on whether the question «Did you see the flash?» is raised or not![7]

Second, let us go back to the first sequence, that is ***E***:

---

[7] Clearly, the Wigner friend paradox is close in spirit to the Schrödinger cat or the von Neumann chain: essentially, they are all variants of the measurement problem.

$$
\left.\begin{array}{ccc}
t_1 & \Rightarrow & t_2 \\
\\
a|1> + b|2> & \Rightarrow & |1>
\end{array}\right\} E
$$

Even if the description is supposed to be relativized to an *E*-observer, RQM inherits from standard QM the state reduction process: as a consequence, RQM shares with standard QM the controversial status of state reduction. For what Rovelli does essentially is to take standard QM at face value and reformulate it by providing any unqualified, observer-independent state with an *observer-index*. Rovelli seems to argue that such shift to a notion of observer-dependent state either removes or dissolves (or contributes to deflate) 'the measurement problem': but it is hard to see how, as the reference to the Wigner friend paradox shows. If we take QM at face value, we are forced to address the puzzling circumstance following from the formal constraints of the theory. Namely, the fact that according to the quantum-mechanical analysis of a typical (idealized) measurement process, the interaction between a measured system and a measuring apparatus induces a superposition of macroscopic states, a superposition that standard QM 'makes sense of' with the aid of the state reduction assumption: attaching observer-relative indexes to the involved states can hardly said to 'remove' the problem of justifying the status of the reduction process. In other words: even if we relativize the description of a measurement process to 'observers' – though in the neutral sense that Rovelli emphasizes, according to which the status of an observer is just like the status of a reference frame in a mechanical description of motion – still we have an observer-dependent description with a state reduction process for which we have no deeper justification. Moreover, in the *E*-description Rovelli as a matter of fact seems to assume that the sequence of events in the measurement *does not include the apparatus as a quantum system*: that's why the arrow goes from $a|1> + b|2>$ to $|1>$, which are states of the system *s* only. This looks like a sort of Bohrian approach: in principle it is not an immediately inconsistent approach *in itself* but, as is well known, it leads to very serious problems in deciding

(i) where the classical/quantum divide is supposed to be located, and (ii) when an interaction is supposed to be a 'measurement'- or a 'non-measurement'-interaction. Moreover, it seems to contradict a key point in the Rovelli approach according to which «All systems are equivalent: nothing *a priori* distinguishes macroscopic systems from quantum systems.» (Rovelli 1996, p. 1644).

If we take seriously the problem arising from a such Bohr-like view, then, we have to include the apparatus – the 'observer' $O$ – into the description even when the sequence $E$ is concerned (just like standard QM). But this leads immediately to the description $E'$

$$\left. \begin{array}{ccc} t_1 & \Rightarrow & t_2 \\ \\ (a|1> + b|2>) \otimes |O\text{-ready}> & \Rightarrow & a|1> \otimes |O1> + b|2> \otimes |O2> \end{array} \right\} E'$$

which is not a different sequence w.r.t. to $E$, but simply the same sequence under the (standard) assumption that $O$ is a quantum system. As to the situation of the observer $P$, in the account of the third person problem there is an ambiguity in claiming that $P$ 'describes' the system $s$-$O$ but *without doing anything*: according to the role that relational QM ascribes to the notion of information, there is no way of acquiring information without interaction (in terms of correlation). For instance

The fact that the pointer variable in $O$ has information about $s$ (has measured **A**) is expressed by the existence of a correlation between the **A** variable of S and the pointer variable of O. The existence of this correlation is a measurable property of the *O-s* state. (Rovelli 1996, p. 1652)

or, as we read in the van Fraassen account of RQM:

The only way in which there can be information for one observer of what has happened to another observer is through a physical measurement by the former on the latter. Communication, i.e. exchange of information, is physical. (van Fraassen 2010, p. 398)

So it seems reasonable to say that RQM inherits from standard QM the ordinary way to describe the kind of interaction that is suitable to account for measurements, namely through the tensor product coupling. But in this case, we couple $H_P$ to ($H_s \otimes H_O$) so as to obtain $H_P \otimes H_s \otimes H_O$, and in turn we obtain the sequence (with the obvious interpretation for the states $|P\text{-ready}>$, $|P1>$ and $|P2>$:

$(a|1>+b|2>) \otimes |O\text{-ready}> \otimes |P\text{-ready}> \quad \Rightarrow \quad a|1> \otimes |O1> \otimes |P1> + b|2> \otimes |O2> \otimes |P2>$

which is completely consistent with *E* and *E′*, provided that now the system under scrutiny is *P+s+O* and no more *s+O*.

To sum up: the "third person problem" is allegedly one of the cornerstones of RQM, and the main argument underlying it is meant to support the shift to a fundamental relativization of states of quantum physical systems to observers. The formulation itself of the problem, however, seems to be based on a basic ambiguity, affecting the very description of a measurement process provided by different observers. If one tries to remove the ambiguity by sharpening the description, the alleged difference between *E* and *E′* appears to evaporate and the motivation for the relativization gets considerably weaker. Moreover, even if we set aside the ambiguity inherent in the formulation of the third person problem, it is hard to see how even a fundamental relativization of states of quantum physical systems to observers in RQM might make sense without further justification of the collapse process, that remains unexplained in the interpretation: hence, the measurement problem *stricto sensu* remains unsolved in the interpretation.

# 4 RQM and the EPR-Bell non-locality issue

Since the original publication in 1935, the EPR correlations and Bell's theorem do not stop to puzzle physicists and philosophers alike. Several attempts have been made in order to deflate the revolutionary impact of these results and, as a matter of fact, RQM sides with them. The more recent version of this deflationary attitude tries to cast the issue in terms of a vague 'realism', a (far from well specified) condition whose conjunction with locality would be the alleged target of Bell's theorem. Thus what is supposed to be the focus of the latter, jointly with the other (obvious) assumption that quantum-mechanical predictions are to preserved, is summarized in the expression *local realism*: the interpretation focusing on local realism would then allow one to preserve locality, by 'discharging' the import of the theorem on the side of "realism". Given the anti-realistic folklore surrounding quantum mechanics anyway since the inception of the theory, this move comes at a relatively little price and contributes to downplay significantly the foundational relevance of Bell's theorem. In this spirit, RQM envisions "the possibility of reading EPR-type experiments as a challenge to Einstein's strong realism, rather than locality." (Smerlak, Rovelli 2007 p. 428).[8]

The first remark to be made, before analyzing how RQM is supposed to implement the above attitude, is that RQM grounds its reading of EPR correlations on an alternative between 'strong realism' and locality which in fact is *mistakenly* assumed to be the core of Bell's theorem. In the RQM approach, the relativization of states and the ensuing lack of observer-independence are taken to realize a suitable weakening of an alleged 'strong realism' but, in fact, such realism does *not* belong to the set of independent assumptions of Bell's theorem (Laudisa 2012, 2017). As is well known, the EPR argument can be formulated as an inference from three conditions to the incompleteness of quantum mechanics: the first is consistency with the statistical predictions of quantum mechanics, the second is the infamous "element-of-physical-reality" condition, whose original formulation in the EPR paper Einstein was dissatisfied with (Howard 1985), and the third is of course locality. It must be stressed that assuming the "element-of-physical-reality" condition is *not* equivalent to assuming the *existence of elements-of-physical-reality* as an autonomous condition: on the contrary, assuming this condition simply amount to require a *criterion* in order for a property of a physical system to be an objective (i.e. measurement-independent) property[9]. The effective existence of such properties is rather *a consequence* of the locality assumption. Since we start with an entangled state of a composite system, in which spin properties of each EPR subsystem are not elements of physical reality and we end dealing with post-measurement states in which such properties *are* indeed elements of physical reality, the only option open to a *local* description of the whole process is that those properties *were already there*, and this is something we *derive* from our assumption that all physical processes involved in the EPR-preparation-and-measurement procedure must be local. But if such existence of elements-of-physical-reality is *not* assumed at the outset, we cannot dismiss anymore the Bell theorem as a nonlocality result simply by charging it with the accusation of smuggling some 'classical realism' – whatever it is – into the description since the beginning. If this is true, then we must acknowledge that what the Bell theorem is about is nonlocality. In this sense, the simple decision to drop observer-independence of states and values of physical quantities can *not* be in itself a first step toward the preservation of locality within the quantum domain, because in this case the lack of observer-independence turns out to be a 'challenge' to a condition which is *not* an assumption of Bell's theorem.

---

[8] Eminent physicists have shared this attitude, such as Nobel laureate Sir Antony J. Leggett: "I believe that the results of the present investigation provide quantitative backing for a point of view which I believe is by now certainly well accepted at the qualitative level, namely that the incompatibility of the predictions of objective local theories with those of quantum mechanics has relatively little to do with locality and much to do with objectivity." (Leggett 2003, p. 1470)

The real challenge is to drop the basic assumption according to which, after the experiment on either given side is performed, its ±1-valued outcome is an observer-independent element of physical reality: it is *this* assumption that holds in any general EPR framework but that in RQM does not hold anymore. Let us consider the point in a specific formulation of the EPR argument in the Bohm version, in which we have a composite quantum system $S_1+S_2$ of a pair of spin-1/2 particles $S_1$ and $S_2$. The composite system is prepared at a time $t_0$ in the singlet state $\Psi_{1+2}$

$$\Psi_{1+2} = 1/\sqrt{2}\,(|1,+>_n |2,->_n - |1,->_n |2,+>_n ),$$

where **n** denotes a generic spatial direction. We take into account the measurements concerning the spin components along given directions, whose possible outcomes are only two (conventionally denoted by '+1' and '−1'). We assume also that the spin measurements on $S_1$ and $S_2$ are performed when $S_1$ and $S_2$ occupy two mutually isolated spacetime regions $R_1$ and $R_2$. According to QM, we know that if the state of $S_1+S_2$ at time $t_0$ is $\Psi_{1+2}$, then the (reduced) states of the subsystems $S_1$ and $S_2$ at time $t_0$ are respectively

$$\rho_1 = 1/2(\mathbf{P}_{|1,+>n} + \mathbf{P}_{|1,->n}), \qquad \rho_2 = 1/2(\mathbf{P}_{|2,+>n} + \mathbf{P}_{|2,->n}) \qquad \textbf{(RS)}$$

---

[9] Maudlin 2014, Laudisa 2017.

('**RS**' stands for '**R**educed **S**tates') so that, for any **n,**

$$\text{Prob}_{\rho_1}\,(\text{spin }_n \text{ of } S_1 = +1) = \text{Prob}_{\rho_1}\,(\text{spin }_n \text{ of } S_1 = -1) = 1/2$$

$$\text{Prob}_{\rho_2}\,(\text{spin }_n \text{ of } S_2 = +1) = \text{Prob}_{\rho_2}\,(\text{spin }_n \text{ of } S_2 = -1) = 1/2$$

Moreover, if we perform at a time $t$ a spin measurument on $S_1$ along **n** with outcome +1 (−1), a spin measurement on $S_2$ along **n** at a time $t' > t$ will give with certainty the outcome − 1 (+1), namely for any **n** ('**AC**' stands for '**A**nti**C**orrelation')

$$\text{Prob}_\Psi\,[(\text{spin }_n \text{ of } S_1 = +1)\ \&\ (\text{spin }_n \text{ of } S_2 = -1)] = \qquad \textbf{(AC)}$$

$$\text{Prob}_\Psi\,[(\text{spin }_n \text{ of } S_1 = -1)\ \&\ (\text{spin }_n \text{ of } S_2 = +1)] = 1.$$

Let us suppose now to perform at time $t_1 > t_0$ a spin measurement on $S_1$ with outcome +1. Therefore, according to (**AC**), a spin measurement on $S_2$ along **n** at a time $t_2 > t_1$ will give with certainty the outcome − 1. Let us suppose now to assume the following condition:

**REALITY** – If, without interacting with a physical system *S*, we can predict with certainty - or with probability 1 - the value **q** of a quantity **Q** pertaining to *S*, then **q** represents an objective property of *S* (denoted by [**q**]).

Then, for $t_2 > t_1$ [**spin** $_n = -1$] represents an objective property of $S_2$. But might the objective property [**spin** $_n = -1$] of $S_2$ have been somehow "created" by the spin measurement on the distant system $S_1$? The answer is negative if we assume the following condition:

**LOCALITY** – No objective property of a physical system *S* can be influenced by operations performed on physical systems that are isolated from *S*.

At this point, **LOCALITY** allows us to state the existence of the objective property [**spin** $_n = -1$] for the system $S_2$ also at a time *t'* such that $t_0 > t' > t_1$. Namely, if we assume that the measurement could not influence the validity of that property at that time, it follows that the property *was holding already at time t'*, a time that *precedes* the measurement performed on the other subsystem. But at time *t'* the state of $S_1 + S_2$ is the singlet state $\Psi_{1+2}$, therefore according to (**RS**) the state of $S_2$ is the reduced state $\rho_2 = 1/2(\mathbf{P}_{|2,+>n} + \mathbf{P}_{|2,->n})$, that prescribes for the property [**spin** $_n = -1$] of $S_2$ only a probability 1/2. Let us consider finally the following condition:

**COMPLETENESS** – Any objective property of a physical system *S* must be represented within the physical theory that is supposed to describe *S*.

It follows that there exist properties of physical systems that, according to the **REALITY** condition are objective – like [**spin**$_n = -1$] for $S_2$ – but that QM does not represent as such: therefore QM is not complete.

RQM attacks the argument in a key point, namely the hypothesis that – according to (**AC**) – the outcome – 1 (+1) that should be yielded with certainty by a spin measurement on $S_2$ along **n** at a time *t' > t*, after a spin measurement on $S_1$ along **n** with outcome +1 (–1) has been performed at a time *t < t'*, is a definite property *also for $S_1$*:

Einstein's argument relies on the strongly realistic hypothesis that *the actual properties of the particles* (*the "real state of affairs"*) *revealed by the detectors are observer-independent*. It is this hypothesis that justifies the ascription of a definite, objective state to each particle, at every instant of the experiment: in Einstein's account, when [the observer] *B* measures the spin of β [our $S_2$], the measured value *instantaneously acquires an objective existence also relative to A* [our $S_1$]. This hypothesis, namely that when *B* measures the spin of β, the measured value instantaneously acquires an objective existence that can be considered absolute, is common to all the analyses that lead to an interpretation of the EPR correlations as a manifestation of non-locality. (Smerlak, Rovelli 2007, pp. 435-436, emphasis in the original)

By the point of view of RQM, if this hypothesis does not hold the EPR argument cannot even be formulated. As a consequence, non-locality for QM would be a non-starter, to the extent that the EPR argument is – via Bell's theorem – the ground for non-locality: but if giving up this hypothesis is a viable option that can be adopted in principle, several points can be raised about the approach to the presuppositions of the EPR argument that this option licenses.

The first is a general, meta-theoretical one. The operational stance implicit by RQM, presented as a refinement of "the original motivations of Bohr and Heisenberg, in order to make full sense of quantum mechanics" (Smerlak, Rovelli 2007, p. 443) is by no means a necessity. A great deal of scientific work in current physics shows that under certain conditions the inaccessibility of certain entities, processes or states need not prevent to reasonably assume their existence and observer-independent definiteness. This point can be supported not only concerning some approaches to the foundations and interpretation of quantum mechanics, such as Bohmian mechanics and many-worlds interpretation, in which the information about, respectively, the exact distribution of particles and the actual perception of what goes on simultaneously at different branches of the wave function is inaccessible *in principle*, but also to other respected areas of research in theoretical physics such as string theory or inflationary cosmological models.[10]

Second, there are approaches which at first sight share with RQM the refusal to ascribe actual properties to the individual EPR subsystems in an observer-independent fashion. For instance, in a paper dating back to 1982 Don Page remarked:

Then why did it appear to EPR that either observable of subsystem 1 may be predicted (without disturbing it) by measuring subsystem 2? Here an ambiguity in the use of «prediction» enters, for in the quantum analysis what is predicted is actually not an observable of subsystem 1 alone but a correlation between subsystem 1 and whatever has measured subsystem 2." (Page 1982, p. 58).

According to Page, therefore, what happens in a typical EPR context when a 'measurement' takes place is just the establishment of a correlation between the apparatus' states and the states of subsystems 1 and 2: starting from the state $\Psi_{1+2} = 1/\sqrt{2} \,(|1,+\rangle_n |2,-\rangle_n - |1,-\rangle_n |2,+\rangle_n)$ and the above mentioned possible states of an observer, i.e. $\{|O\text{-}ready\rangle, |O+\rangle, |O-\rangle\}$, where in the obvious interpretation $|O\text{-}ready\rangle$ represents the state of the observer $O$ prior to the measurement, $|O+\rangle$ represents the state of the observer $O$ finding the measurement value + and $|O-\rangle$ represents the state of the observer $O$ finding the measurement value –, a correlation

$$\Psi_{O+1+2} = 1/\sqrt{2} \,(|O+\rangle|1,+\rangle_n |O-\rangle|2,-\rangle_n |O1\rangle - |O-\rangle|1,-\rangle_n |O+\rangle|2,+\rangle_n)$$

gets established, from which the reduced states of $O$, 1 and 2 respectively

$$\rho_O = 1/2 \,(\mathbf{P}_{|O+\rangle} + \mathbf{P}_{|O-\rangle})$$

$$\rho_1 = 1/2 \,(\mathbf{P}_{|1,+\rangle_n} + \mathbf{P}_{|1,-\rangle_n}),$$

$$\rho_2 = 1/2 \,(\mathbf{P}_{|2,+\rangle_n} + \mathbf{P}_{|2,-\rangle_n}),$$

can be derived. Since the Page approach is a *no-collapse* account of the EPR correlations, in which the latter are exclusively properties of the joint system, clearly the above correlation need

---

[10] It may suffice to recall the hot debates on the hypothesis of the *multiverse* (Ellis 2011) or on the implications for the very definition and soundness of scientific methodology in the area of string theory (Dawid 2013).

not single out a unique value since the apparatus' states simply get correlated each with one component of the mixture:

Thus we see that the EPR predictions always involves a comparison of subsystem 1 and 2 or of apparatuses which have measured them. The prediction may never be checked by a measurement of either subsystem alone. Hence EPR are incorrect in ascribing reality to the precise values of observables of subsystem 1. These observables do not have precise values, but only expectation values in the state considered. The physical reality that can be predicted, namely the correspondence between the two measurement results when compared, does have a counterpart in the density operator (or wave function, for a pure state) of the entire composite system. (Page 1982, p. 59) [11]

These approaches, however, may afford to establish such a conclusion because they explicitly drop the collapse postulate, an assumption that – although in a relativized sense – RQM *does* preserve (remember the sequence *E* in the discussion of the "third person" problem).

The final point concerns the implications for RQM of a distinction that was introduced long time ago in the context of stochastic hidden variable models of quantum mechanics (Jarrett 1984): a distinction between two conditions, called respectively *parameter independence* and *outcome independence*, through which a 'peaceful coexistence' between quantum mechanics and relativity theory about nonlocality might have been achieved (Shimony 1984). Let us employ the symbol λ to denote all parameters useful to characterize the complete specification of the state of an individual physical system (the presentation follows Ghirardi *et al.* 1993). In a standard EPR-Bohm-like situation, the expression

$$P_\lambda^{LR}(x, y; \mathbf{n}, \mathbf{m}) \qquad (1)$$

denotes the joint probability of getting the outcome $x$ ($x = \pm 1$) in a measurement of the spin component along $\mathbf{n}$ at the left (L), and $y$ ($y = \pm 1$) in a measurement of the spin component along $\mathbf{m}$ at the right (R) wing of the apparatus. We assume that the experimenter at L can make a free-will choice of the direction $\mathbf{n}$ and similarly for the experimenter at R and the direction $\mathbf{m}$. Both

---

[11] Tipler 2014 argues along similar lines, although in a more explicitly Everettian vein.

experimenters can also choose not to perform the measurement. Bell's locality assumption can be expressed as

$$P_\lambda^{LR}(x, y; \mathbf{n}, \mathbf{m}) = P_\lambda^{L}(x; \mathbf{n}, *) \, P_\lambda^{R}(y; *, \mathbf{m}) \qquad (2)$$

where the symbol * in the probabilities at the r.h.s denotes that the corresponding measurement is not performed. Jarrett has shown that condition (2) is equivalent to the conjunction of two logically independent conditions, namely (Jarrett 1984)

$$P_\lambda^{L}(x; \mathbf{n}, \mathbf{m}) = P_\lambda^{L}(x; \mathbf{n}, *) \qquad (3a)$$

$$P_\lambda^{R}(y; \mathbf{n}, \mathbf{m}) = P_\lambda^{R}(y; *, \mathbf{m})$$

and

$$P_\lambda^{LR}(x, y; \mathbf{n}, \mathbf{m}) = P_\lambda^{L}(x; \mathbf{n}, \mathbf{m}) \, P_\lambda^{R}(y; \mathbf{n}, \mathbf{m}) \qquad (3b)$$

Conditions (3a) – referred to as *parameter independence* (PI) – jointly express the requirement that the probability of getting a result at L (R) is independent from the setting chosen at R (L), while condition (3b) – referred to as *outcome independence* (OI) – expresses the requirement that the probability of an outcome at one wing does not depend on the outcome which is obtained at the other wing. Now it can be shown that standard quantum mechanics *does* violate one of the above independence conditions, namely outcome independence. For in the standard EPR case when $\lambda$ is the singlet state $\Psi$, if we choose $\mathbf{n} = \mathbf{m}$ we get

$$P_\Psi^{LR}(1, -1; \mathbf{n}, \mathbf{n}) = P_\Psi^{LR}(-1, 1; \mathbf{n}, \mathbf{n}) = 1/2 \qquad (4)$$

$$P_\Psi^{LR}(1, 1; \mathbf{n}, \mathbf{n}) = P_\Psi^{LR}(-1, -1; \mathbf{n}, \mathbf{n}) = 0$$

but for any $x, y$

$$P_\Psi^{L}(x; \mathbf{n}, \mathbf{n}) \, P_\Psi^{R}(y; \mathbf{n}, \mathbf{n}) = 1/4, \qquad (5)$$

a result that shows the quantum mechanical violation of outcome independence for certain choices of parameters[12]. Now, according to the basic tenets of RQM, the relativization to specific states of reference quantum systems is assumed to coexist with the probabilistic statements of standard QM, that RQM is supposed to preserve completely. This is taken to fulfil two different requirements. The first is to satisfy the obvious need of RQM to agree on quantum predictions, since RQM aims at diverging from good old Copenhagen QM as little as possible. The second is to give a further expression to the analogy with relativity that Rovelli hinted at when proposing RQM itself for the first time: the coexistence in RQM between the relational nature of all physical statements of state attribution and the observer-*in*dependence of probabilistic statements is supposed to work along the lines of the coexistence – in standard QM – between the frame-dependent character of post-collapse states of quantum systems on one hand and the the frame-*in*dependence of probabilistic statements, whose predictions hold independently of the Lorentz frame that can be used (Gambini, Porto 2002). But if this is true, RQM appears then to be forced to accept *some* form of non-locality in quantum phenomena, namely the non-local sort of dependence that shows up in the above mentioned violation of outcome-dependence, which is a *probabilistic* condition.

## 5  Conclusions

In the wide area of the interpretations of QM, the relational approach first developed by Carlo Rovelli attempts to jointly address some of the most pressing problems of the ordinary formulation of the theory: the shifty split between classical and quantum regimes, the ensuing measurement problem, the prospects of a puzzling, 'non-local' picture of the quantum world. Moreover, this approach aims at obtaining a satisfactory interpretation of the theory – that is, a local and measurement-determinate account of quantum phenomena – with minimal modifications with respect to the Copenhagen-style view of QM, summarized in the decision to drop the notion of observer-independent state of a physical system. In the above pages I have attempted to show that we have reasons to be much less optimistic toward the prospects of RQM: a lot of work needs to be done before RQM may aspire to become a satisfactory interpretational framework for the main foundational issues in QM.

---

[12] The peaceful coexistence thesis, recalled above, is grounded on the fact that standard quantum mechanics – although violating OI – *does* satisfy PI (Ghirardi *et al.* 1980). As a consequence, quantum-mechanical non-locality would not be so harmful: the outcomes are somehow non-locally affecting each other and this seems to threaten the prescriptions of special relativity, but such outomes are uncontrollable and thus we cannot exploit them to produce any robust action-at-a-distance. The effectiveness of the PI/OI distinction in carrying the burden of such an ambitious coexistence has been often questioned: see Maudlin 1994, 2011³, pp. 85-90, for a critical analysis.